\documentclass[12pt]{article}
\usepackage{graphicx}
\usepackage{amsmath}
\usepackage{amsbsy}
\usepackage{amsfonts}
\usepackage{amsthm}
\begin{document}

\title{Q-stars in scalar-tensor gravitational theories}

\author{Athanasios Prikas}

\date{}

\maketitle

Physics Department, National Technical University, Zografou Campus,
157 80 Athens, Greece, tel. +302107722991.\footnote{e-mail:
aprikas@central.ntua.gr}

\begin{abstract}
We study q-stars in Brans-Dicke gravitational theory. We find that
when the Brans-Dicke constant, $\omega_{\textrm{BD}}$, tends to
infinity, the results of General Relativity are reproduced. For
other values of $\omega_{\textrm{BD}}$, the particle number, mass
and radius of the star and the absolute value of the matter field
are a few percent larger than in the case of General Relativity.
We also investigate the general scalar-tensor gravitational
theory and find that the star parameters are a few percent larger
than in the case of General Relativity.
\end{abstract}

PACS number(s): 11.27.+d, 04.40.-b

\newpage

\section{Introduction}

Interesting alternative gravitational theories are the
scalar-tensor gravitational theories, which appeared at the
original paper of Brans and Dicke \cite{bransdicke}, where the
Newtonian constant $G$ was replaced by a scalar field
$\phi_{\textrm{BD}}$, and the total action contained kinetic terms
for the new field times an $\omega_{\textrm{BD}}$ quantity.
$\omega_{\textrm{BD}}$ was regarded as a constant in the original
paper. The theory was generalized in a series of papers,
\cite{scalartensor1,scalartensor2}, mainly in the direction of
replacing the constant $\omega_{\textrm{BD}}$ with a function of
the Brans-Dicke (BD) scalar field.

Boson stars appeared in the literature as stable field
configurations of massive scalar matter with a global $U(1)$
symmetry, coupled to gravity
\cite{boson-stars-old1,boson-stars-old2}, now known as ``mini"
boson stars due to their small relative magnitude. Other works
took into account self interactions
\cite{boson-stars-new,boson-stars-quartic} or the case of local
symmetry \cite{boson-stars-charged}. Their common feature is that
gravity plays the role of the non-linear interaction that
stabilizes the star against decay into free particles.

When the scalar potential is of a special type, admitting stable
non-topological soliton solutions in the absence of gravity, the
soliton stars appear as relativistic generalizations of the above
solitons. The so called ``large" soliton stars, with radius of order
of lightyears, discussed analytically in a series of papers by
Friedberg, Lee and Pang,
\cite{large-soliton-stars1,large-soliton-stars2,large-soliton-stars3}.
Another class of soliton stars appeared as a generalization of
q-balls. Q-balls are non-topological solitons in Lagrangians with a
global $U(1)$ symmetry, \cite{qballs-initial}, or a local one,
\cite{qballs-charged}, or a global $SU(3)$ or $SO(3)$ symmetry,
\cite{qballs-nonabelian}. Q-balls are supposed to appear in the flat
directions of the superpotential in supersymmetric extensions of
Standard Model, \cite{qballs-mssm1,qballs-mssm2}, and play a special
role in the baryogenesis, \cite{qballs-baryogenesis}. Q-stars are
relativistic extensions of q-balls, with one or two scalar fields
and a global, \cite{qstars-global}, or local, \cite{qstars-charged},
$U(1)$ symmetry, non-abelian symmetry, \cite{qstars-nonabelian}, or
with fermions and a scalar field, \cite{qstars-fermion} in
asymptotically flat or anti de Sitter spacetime, \cite{qstars-AdS}.
Any type of the bosonic stars may offer a solution to the problem of
Dark Matter, when the q-stars have the additional feature to be of
the same order of magnitude as neutron stars and, generally, as
other stellar objects.

Within the BD gravitational framework, Gunderson and Jensen
investigated the coupling of a scalar field with quartic
self-interactions with the metric and the BD scalar field,
$\phi_{\textrm{BD}}$, \cite{bosonstar-bransdicke1}. The properties
of boson stars within this framework have been extensively studied
in a series of papers
\cite{bosonstar-bransdicke2,bosonstar-bransdicke3,bosonstar-bransdicke4}.
Their results generalized in scalar-tensor gravitational theories,
where $\omega_{\textrm{BD}}$ is no more a constant, but a
function of the BD field
\cite{bosonstar-scalartensor1,bosonstar-scalartensor2}. The case
of charged boson-stars in a scalar-tensor gravitational theory
has been analyzed in \cite{bosonstar-scalartensor-charged}.

In the present article we follow the work of
\cite{bosonstar-bransdicke1}-\cite{bosonstar-scalartensor-charged}.
Our aim is to study the formation of non-topological soliton stars
in the context of BD or general scalar-tensor gravitational theory,
their stability with respect to fission into free particles and to
gravitational collapse, and the influence of $\omega_{\textrm{BD}}$
in the star parameters. We also compare our results with those
obtained in the framework of General Relativity.

\section{Q-stars with one scalar field}

We consider a static, spherically symmetric metric:
\begin{equation}\label{2.1}
ds^2=-e^{\nu}dt^2+e^{\lambda}d\rho^2+\rho^2d\Omega^2\ ,
\end{equation}
with $g_{tt}=-e^{\nu}$. In order to realize such a spacetime, we
regard both the matter and the BD fields as spherically symmetric
and the former with an harmonic time dependence, assuring minimum
energy for the matter field, and the latter time independent. If
$\phi$ is the matter field and $\phi_{\textrm{BD}}$ the BD field,
we write the action for the BD theory:
\begin{align}\label{2.2}
S=\frac{1}{16\pi}&\int
d^4x\sqrt{-g}\left(\phi_{\textrm{BD}}R-\omega_{\textrm{BD}}g^{\mu\nu}
\frac{\partial_{\mu}\phi_{\textrm{BD}}\partial_{\nu}\phi_{\textrm{BD}}}{\phi_{\textrm{BD}}}
\right) \nonumber
\\ + &\int d^4x\sqrt{-g}\mathcal{L}_{\textrm{matter}}\ ,
\end{align}
with:
\begin{equation}\label{2.3}
\mathcal{L}_{\textrm{matter}}=(\partial_{\mu}\phi)^{\ast}(\partial^{\nu}\phi)-U
\ ,
\end{equation}
and $\omega_{\textrm{BD}}$ a constant in BD gravity and a certain
function of the $\phi_{\textrm{BD}}$ field in a generalized
scalar-tensor gravitational theory, which we will discuss later.

Varying the action with respect to the metric and scalar fields
we obtain the Einstein and Lagrange equations respectively, as
follows:
\begin{align}\label{2.4}
G_{\mu\nu}=\frac{8\pi}{\phi_{\textrm{BD}}}T_{\mu\nu}+
\frac{1}{\phi_{\textrm{BD}}}({\phi_{\textrm{BD}}}_{,\mu;\nu}-g_{\mu\nu}
{\phi_{\textrm{BD}}}_{;\lambda}^{\hspace{1em};\lambda} )
\nonumber\\ +
\frac{\omega_{\textrm{BD}}}{\phi_{\textrm{BD}}^2}\left(
\partial_{\mu}\phi_{\textrm{BD}}\partial_{\nu}\phi_{\textrm{BD}}-
\frac{1}{2}g_{\mu\nu}\partial_{\lambda}\phi_{\textrm{BD}}
\partial^{\lambda}\phi_{\textrm{BD}}\right)\ ,
\end{align}
\begin{equation}\label{2.5}
\phi_{;\lambda}^{\hspace{1em};\lambda}-\frac{dU}{d|\phi|^2}\phi=0
\ ,
\end{equation}
\begin{equation}\label{2.6}
\frac{2\omega_{\textrm{BD}}}{\phi_{\textrm{BD}}}
{\phi_{\textrm{BD}}}_{;\lambda}^{\hspace{1em};\lambda}-
\omega_{\textrm{BD}}
\frac{\partial^{\lambda}\phi_{\textrm{BD}}\partial_{\lambda}\phi_{\textrm{BD}}}
{\phi_{\textrm{BD}}^2}+R=0 \ .
\end{equation}
$G_{\mu\nu}$ is the Einstein tensor, $T_{\mu\nu}$ the energy
momentum tensor for the matter field given by:
\begin{equation}\label{2.7}
T_{\mu\nu}={({\partial}_{\mu}\phi)}^{\ast}({\partial}_{\nu}\phi)+
({\partial}_{\mu}\phi){({\partial}_{\nu}\phi)}^{\ast}
-g_{\mu\nu}[g^{\alpha\beta}{({\partial}_{\alpha}\phi)}^{\ast}({\partial}_{\beta}\phi)]
-g_{\mu\nu}U\
\end{equation}
and $R$ is the scalar curvature. Tracing eq. \ref{2.4} we take:
$$-\frac{8\pi}{\phi_{\textrm{BD}}}T-\frac{\omega_{\textrm{BD}}}{\phi_{\textrm{BD}}^2}
\partial^{\lambda}\phi_{\textrm{BD}}\partial_{\lambda}\phi_{\textrm{BD}}
+\frac{3}{\phi_{\textrm{BD}}}{\phi_{\textrm{BD}}}^{\hspace{1em};\lambda}_{;\lambda}=R\
.$$ with $T$ the trace of the energy-momentum tensor. Substituting
in eq. \ref{2.6} we find:
\begin{equation}\label{2.8}
{\phi_{\textrm{BD}}}^{\hspace{1em};\lambda}_{;\lambda}
=\frac{8\pi}{2\omega_{\textrm{BD}}+3}T \ .
\end{equation}
The above results hold true for every case of bosonic, spherically
symmetric, static field configurations coupled to BD gravity.

We will now insert the q-soliton ansatz writing:
\begin{equation}\label{2.9}
\phi(\vec{\rho},t)=\sigma(\rho)e^{-\imath\omega t} \ .
\end{equation}
with $\omega$ the frequency with which the q-soliton rotates
within its internal $U(1)$ space. The Lagrange equation for the
$\phi$ field is:
\begin{equation}\label{2.10}
\sigma''+[2/\rho+(1/2)(\nu'-\lambda')]\sigma'+
e^{\lambda}\omega^2e^{-\nu}\sigma-e^{\lambda}\frac{dU}{d\sigma^2}\sigma=0\
.
\end{equation}
We define:
\begin{equation}\label{2.11}
A=e^{-\lambda}\ , \hspace{1em} B=e^{-\nu}\ ,
\end{equation}
\begin{equation}\label{2.12}
\begin{split}
W\equiv e^{-\nu}{\left(\frac{\partial\phi}{\partial
t}\right)}^{\ast}\left(\frac{\partial\phi}{\partial t}\right)=
e^{-\nu}{\omega}^2{\sigma}^2\ , \\ V\equiv
e^{-\lambda}{\left(\frac{\partial\phi}{\partial\rho}\right)}^{\ast}
\left(\frac{\partial\phi}{\partial\rho}\right)=
e^{-\lambda}{\sigma'}^2
\end{split}
\end{equation}
and rescale:
\begin{equation}\label{2.13}
\begin{split}
\tilde{\rho}=\rho m\ , \hspace{1em} \tilde{\omega}&=\omega/m\ ,
\hspace{1em} \tilde{\phi}=\phi/m \ , \\ \widetilde{U}=U/m^4\ ,
\hspace{1em} \widetilde{W}&=W/m^4\ , \hspace{1em}
\widetilde{V}=V/m^4\ .
\end{split}
\end{equation}
In roughly approximation, gravity becomes important when $R\sim
G\mathcal{M}(R)$, where $R$ is defined as the radius, within
which the matter field differs from zero, and $\mathcal{M}(R)$ is
the mass trapped within this area. For the case of q-solitons the
eigen-frequency is of the same order of magnitude as the mass and
the absolute value of the scalar field: $\omega\sim m\sim\sigma$.
If $\mathcal{E}$ is the energy density, then: $\mathcal{E}\sim
U\sim m^4$. Remembering that
$\mathcal{M}(R)\equiv\int_0^R\mathcal{E}d^3\rho$, we find that
for a q-star:
$$ R\sim \epsilon^{-1}\ , \hspace{1em} \epsilon\equiv\sqrt{8\pi
Gm^2}\ ,
$$
so if we redefine:
\begin{equation}\label{2.14}
\tilde{r}=\epsilon\tilde{\rho}\ ,
\end{equation}
we expect $\tilde{r}\sim1$. We also use a suitable rescaled
potential, admitting q-ball type solutions in the absence of
gravity, namely:
\begin{equation}\label{2.15}
\widetilde{U}=|\tilde{\phi}|^2\left(1-|\tilde{\phi}|^2+\frac{1}{3}|\tilde{\phi}|^4\right)
=\tilde{\sigma}^2\left(1-\tilde{\sigma}^2+\frac{1}{3}\tilde{\sigma}^4\right)\
.
\end{equation}

\begin{figure}
\centering
\includegraphics{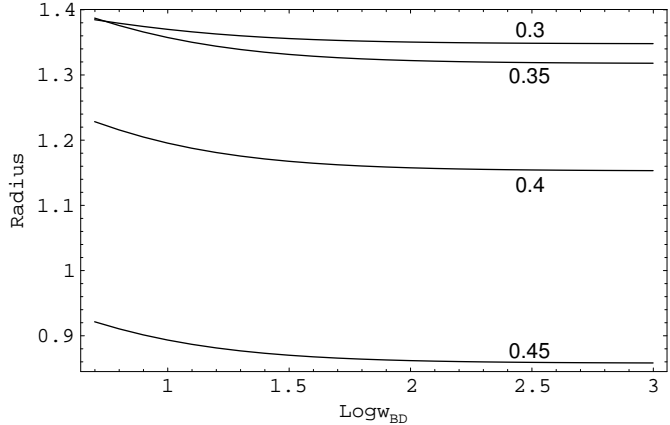}
\caption{The radius of a q-star as a function of
$\omega_{\textrm{BD}}$ for four different values of $\omega$ in a
BD gravitational theory. $\omega$ gives a measure of the gravity
strength at the surface, as it is related with the metrics at the
surface of the star, through eq. \ref{2.19}.} \label{figure1}
\end{figure}

\begin{figure}
\centering
\includegraphics{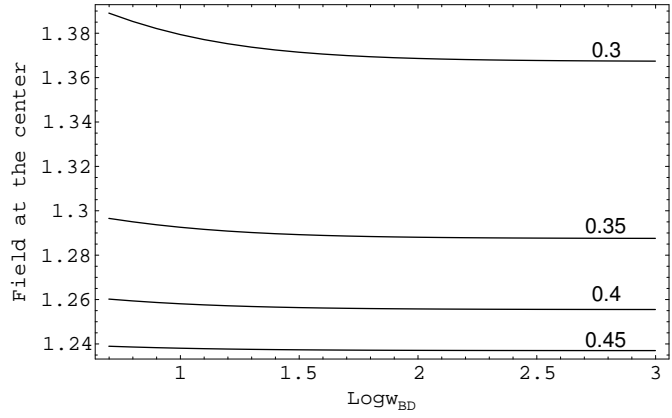}
\caption{The value of the scalar field $\sigma$ at the center of
a q-star as a function of $\omega_{\textrm{BD}}$ for four
different values of $\omega$ in a BD gravitational theory.}
\label{figure2}
\end{figure}

\begin{figure}
\centering
\includegraphics{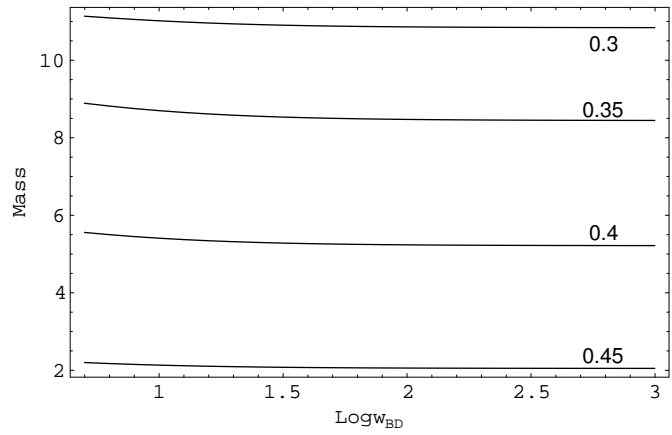}
\caption{The mass of a q-star as a function of
$\omega_{\textrm{BD}}$ for four different values of $\omega$ in a
BD gravitational theory.} \label{figure3}
\end{figure}

\begin{figure}
\centering
\includegraphics{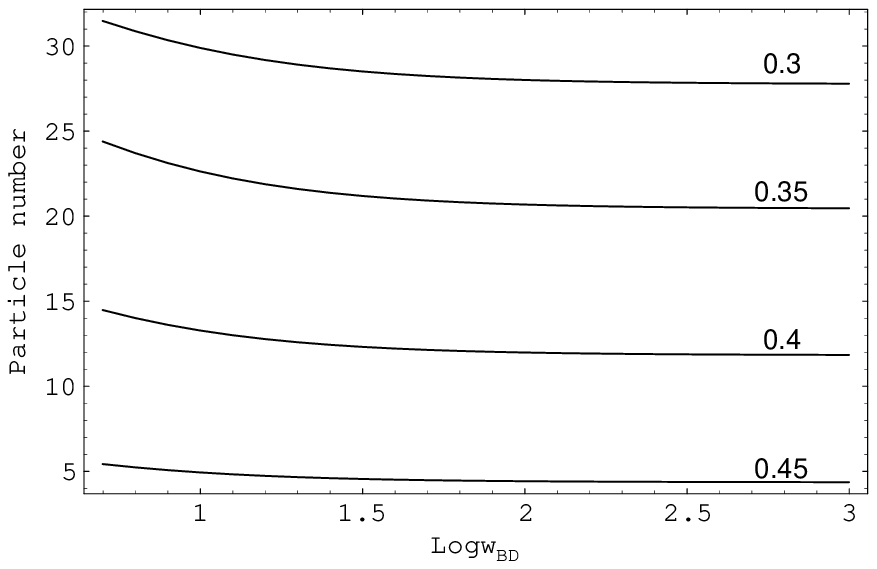}
\caption{The particle number of a q-star as a function of
$\omega_{\textrm{BD}}$ for four different values of $\omega$ in a
BD gravitational theory. From figures
\ref{figure1}-\ref{figure4}, we see that small $\omega$ means a
star with larger radius, mass and particle number.}
\label{figure4}
\end{figure}

Dropping form now on the tildes, we will use the Lagrange
equation to find an analytical solution for the matter field.
$\epsilon$ is a very small quantity for $m$ of the order of some
(hundreds) $GeV$, so ignoring the $O(\epsilon)$ terms from the
Lagrange equation, we find:
\begin{equation}\label{2.16}
\sigma^2=1+\omega B^{1/2}\ ,\hspace{1em}
U=\frac{1}{3}(1+\omega^3B^{3/2})\ .
\end{equation}
The surface is determined by the star radius. The radius of the
solitonic configuration is defined as the radius within which the
\emph{matter} Lagrangian differs from zero. Outside this radius
the matter Lagrangian is zero, when the BD Lagrangian \emph{may
not necessarily be}. The surface width is of order of $m^{-1}$.
Within this, the matter field varies very rapidly from a $\sigma$
value at the inner edge of the surface, to zero at the outer, but
the metric fields vary very slowly. So, dropping from the Lagrange
equation the $O(\epsilon)$ terms we take:
\begin{equation}\label{2.17}
\frac{\delta(W-U-V)}{\delta\sigma}=0\ .
\end{equation}
The above equation can be straightforward integrated and, because
all energy quantities are zero at the outer edge of the surface,
the result gives the following equation, holding true only within
the surface:
\begin{equation}\label{2.18}
V+W-U=0\ .
\end{equation}
At the inner edge of the surface $\sigma'$ is zero in order to
match the interior with the surface solution. So, at the inner
edge of the surface the equality $W=U$ together with eq.
\ref{2.16} gives:
\begin{equation}\label{2.19}
\omega=\frac{A_{\textrm{sur}}^{1/2}}{2}=\frac{B_{\textrm{sur}}^{-1/2}}{2}\
.
\end{equation}
Eq. \ref{2.19} is the eigenvalue equation for the frequency of the
q-star, revealing the relation between a feature of the star and
the spacetime curvature.

Redefining:
\begin{equation}\label{2.20}
\Phi_{\textrm{BD}}=\frac{2\omega_{\textrm{BD}}+3}{2\omega_{\textrm{BD}}+4}
G \phi_{\textrm{BD}}\ ,
\end{equation}
and dropping the $O(\epsilon)$ terms, the Lagrange equation for
the BD field and the Einstein equations take the form
respectively:
\begin{equation}\label{2.21}
A\left[\frac{d^2\Phi_{\textrm{BD}}}{dr^2}+
\left(\frac{2}{r}+\frac{1}{2A}\frac{dA}{dr}-\frac{1}{2B}\frac{dB}{dr}\right)
\frac{d\Phi_{\textrm{BD}}}{dr}\right]=
\frac{2W-4U}{2\omega_{\textrm{BD}}+4}\ ,
\end{equation}
\begin{align}\label{2.22}
\frac{A-1}{r^2}+\frac{1}{r}\frac{dA}{dr}=\frac{2\omega_{\textrm{BD}}+3}
{(2\omega_{\textrm{BD}}+4)\Phi_{\textrm{BD}}}
\left(-W-U-\frac{2W-4U}{2\omega_{\textrm{BD}}+3}\right) \nonumber
\\ -\frac{\omega_{\textrm{BD}}A}{2\Phi_{\textrm{BD}}^2}
\left(\frac{d\Phi_{\textrm{BD}}}{dr}\right)^2-
\frac{A}{2\Phi_{\textrm{BD}}B}\frac{dB}{dr}\frac{d\Phi_{\textrm{BD}}}{dr}\
,
\end{align}
\begin{align}\label{2.23}
\frac{A-1}{r^2}-\frac{A}{B}\frac{1}{r}\frac{dB}{dr}=
\frac{2\omega_{\textrm{BD}}+3}{(2\omega_{\textrm{BD}}+4)\Phi_{\textrm{BD}}}
\left(W-U-\frac{2W-4U}{2\omega_{\textrm{BD}}+3}\right) \nonumber
\\ +\frac{\omega_{\textrm{BD}}A}{2\Phi_{\textrm{BD}}^2}
\left(\frac{d\Phi_{\textrm{BD}}}{dr}\right)^2
+\frac{A}{\Phi_{\textrm{BD}}}\left(\frac{d^2\Phi_{\textrm{BD}}}{dr^2}+
\frac{1}{2A}\frac{dA}{dr}\frac{d\Phi_{\textrm{BD}}}{dr}\right)\ ,
\end{align}
with boundary conditions:
\begin{equation}\label{2.24}
A(0)=1\ , \hspace{1em} A(\infty)=1/B(\infty)=1\ , \hspace{1em}
\Phi_{\textrm{BD}}'=0\ , \hspace{1em}
\Phi_{\textrm{BD}}(\infty)=1\ ,
\end{equation}
where the first condition reflects the freedom to define the
$g_{\rho\rho}$ metric at least locally when the second arises from
the flatness of the spacetime. One may alternatively, instead of
eq. \ref{2.22}, use the relation resulting from the Schwarzschild
formula: $A(\rho)=1-\frac{2G\mathcal{M}(\rho)}{\rho}$, which can
be written with our rescalings as:
\begin{equation}\label{2.25}
A=1-\frac{\mathcal{M}(r)}{4\pi r}\ .
\end{equation}
With the new variable eq. \ref{2.22} takes the form:
\begin{align}\label{2.26}
\frac{1}{4\pi r^2} \frac{d\mathcal{M}}{dr}
=\frac{2\omega_{\textrm{BD}}+3}{(2\omega_{\textrm{BD}}+4)\Phi_{\textrm{BD}}}
\left(-W-U-\frac{2W-4U}{2\omega_{\textrm{BD}}+3}\right)- \nonumber \\
\frac{\omega_{\textrm{BD}}\left(1-\frac{\mathcal{M}}{4\pi
r}\right)}{2\Phi_{\textrm{BD}}^2}\left(\frac{d\Phi_{\textrm{BD}}}{dr}\right)^2
-\frac{1-\frac{\mathcal{M}}{4\pi
r}}{2\Phi_{\textrm{BD}}B}\frac{dB}{dr}\frac{d\Phi_{\textrm{BD}}}{dr}\
,
\end{align}
with $\mathcal{M}(0)=0$ which reflects the absence of anomalies
at the center of the star.

The stability of the star results from a conserved Noether charge.
There is a Noether current due to the global $U(1)$ symmetry
defined as:
\begin{equation}\label{2.27}
j^{\mu}=\sqrt{-g}g^{\mu\nu}\imath(\phi^{\ast}\partial_{\nu}\phi-
\phi\partial_{\nu}\phi^{\ast})
\end{equation}
and a conserved Noether charge defined as:
\begin{equation}\label{2.28}
Q=\int d^3xj^0=8\pi\int drr^2\omega\sigma^2\sqrt{B/A}\ .
\end{equation}

In our figures $R$ is in $(8\pi Gm^4)^{-1/2}$ units, the total
mass in $(8\pi G)^{-3}m^{-2}$ units and the charge in $(8\pi
Gm)^{-3}$ units. The total charge equals to the particle number if
every single particle is assigned with a unity ``baryon" number.
The particle number also equals to the total energy of the free
particles as their mass is taken to be unity. So, when the
particle number exceeds the total mass, the star decays into free
particles as the energetically favorable case. All the field
configurations depicted in our figures are stable. An
experimental lower limit for $\omega_{\textrm{BD}}$ is $500$,
\cite{bosonstar-bransdicke1}. The results obtained in the BD
context coincide with general relativity when
$\omega_{\textrm{BD}}\rightarrow\infty$. We investigate the phase
space of the star with $\omega_{\textrm{BD}}$ varying between $5$
and $1000$, following the works of
\cite{bosonstar-bransdicke1}-\cite{bosonstar-scalartensor-charged},
so as to explore thoroughly the influence of
$\omega_{\textrm{BD}}$ in the features of the star. As a general
result we find that the star parameters, mass, particle number,
radius and absolute value of the scalar field at the center of
the star, increase when $\omega_{\textrm{BD}}$ decreases.

\section{General scalar-tensor theory}

In the original BD gravitational theory $\omega_{\textrm{BD}}$ is
a constant. In a more general theory it may be regarded as a
function, usually of the BD field. We will use one of the forms
that Barrow and Parsons \cite{scalartensor-cosmological}
investigated in a cosmological framework namely:
\begin{equation}\label{5.1}
2\omega_{\textrm{BD}}+3=\omega_0\phi_{\textrm{BD}}^n\ ,
\end{equation}
with $\omega_0$ and $n$ constants. This form for
$\omega_{\textrm{BD}}$ gives an analytical solution,
\cite{scalartensor-cosm-analytical}, for the metrics within the
above mentioned cosmological framework. The Lagrange equation for
the BD field is:
\begin{equation}\label{5.2}
{\phi_{\textrm{BD}}}^{\hspace{1em};\lambda}_{;\lambda}=\frac{1}{2\omega_{\textrm{BD}}+3}
\left(8\pi
T-\frac{d\omega_{\textrm{BD}}}{d\phi_{\textrm{BD}}}
{\phi_{\textrm{BD}}}^{,\rho}{\phi_{\textrm{BD}}}_{,\rho}\right)\
,
\end{equation}
If we rescale:
\begin{equation}\label{5.3}
\tilde{\omega}_0=\left(\frac{2\omega_{\textrm{BD}}+3}{2\omega_{\textrm{BD}}+4}\right)^n
G^n\omega_0\ ,
\end{equation}
and the other quantities as in \ref{2.13}-\ref{2.14} eqs. and
drop the tildes and the $O(\epsilon)$ quantities we take for the
Einstein and the Lagrange equation for the BD field:
\begin{align}\label{5.4}
G_t^{\
t}=\frac{\omega_0}{\omega_0\Phi_{\textrm{BD}}+1}\left[-W-U-\frac{1}
{\omega_0\Phi_{\textrm{BD}}} \right.\times\nonumber\\ \left.
\left(2W-4U-\frac{A\Phi_{\textrm{BD}}'^2}{2}
\frac{\omega_0\Phi_{\textrm{BD}}+1}{\Phi_{\textrm{BD}}}\right)\right]
\nonumber\\
-\frac{\omega_0\Phi_{\textrm{BD}}-3}{2}\frac{A\Phi_{\textrm{BD}}'^2}{2\Phi_{\textrm{BD}}^2}
-\frac{AB'\Phi_{\textrm{BD}}'}{2\Phi_{\textrm{BD}}B}\ ,
\end{align}
\begin{align}\label{5.5}
G_r^{\
r}=\frac{\omega_0}{\omega_0\Phi_{\textrm{BD}}+1}\left[W-U-\frac{1}
{\omega_0\Phi_{\textrm{BD}}} \right.\times\nonumber\\ \left.
\left(2W-4U-\frac{A\Phi_{\textrm{BD}}'^2}{2}
\frac{\omega_0\Phi_{\textrm{BD}}+1}{\Phi_{\textrm{BD}}}\right)\right]
\nonumber\\+\frac{\omega_0\Phi_{\textrm{BD}}-3}{2}\frac{A\Phi_{\textrm{BD}}'^2}
{2\Phi_{\textrm{BD}}^2}+\frac{A\Phi_{\textrm{BD}}''}{\Phi_{\textrm{BD}}}+
\frac{A'\Phi_{\textrm{BD}}'}{2\Phi_{\textrm{BD}}}\ ,
\end{align}
\begin{align}\label{5.6}
A\left[\Phi_{\textrm{BD}}''+\left(\frac{2}{r}+\frac{A'}{2A}-\frac{B'}{2B}\right)
\Phi_{\textrm{BD}}'\right]=\nonumber\\
\frac{1}{\omega_0\Phi_{\textrm{BD}}+1}
\left[2W-4U-\frac{A\Phi_{\textrm{BD}}'^2}{2}\frac{\omega_0\Phi_{\textrm{BD}}+1}
{\Phi_{\textrm{BD}}}\right]\ .
\end{align}

\begin{figure}
\centering
\includegraphics{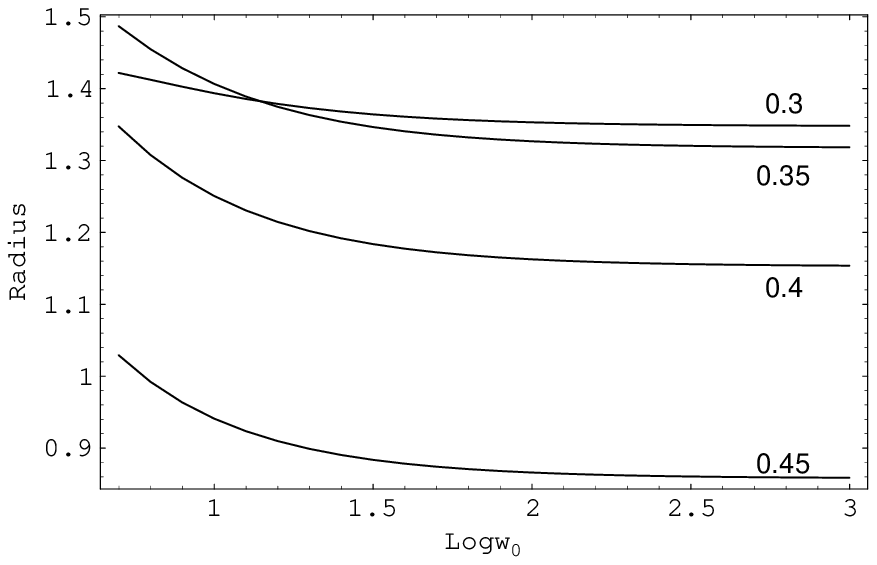}
\caption{The radius of a q-star as a function of $\omega_0$ for
four different values of $\omega$ in a scalar-tensor
gravitational theory with $n=1$.} \label{figure13}
\end{figure}

\begin{figure}
\centering
\includegraphics{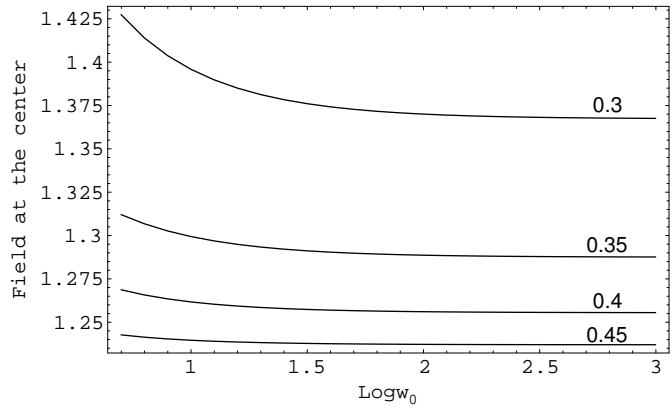}
\caption{The value of the scalar field $\sigma$ at the center of
a q-star as a function of $\omega_0$ for four different values of
$\omega$ in a scalar-tensor gravitational theory with $n=1$.}
\label{figure14}
\end{figure}

\begin{figure}
\centering
\includegraphics{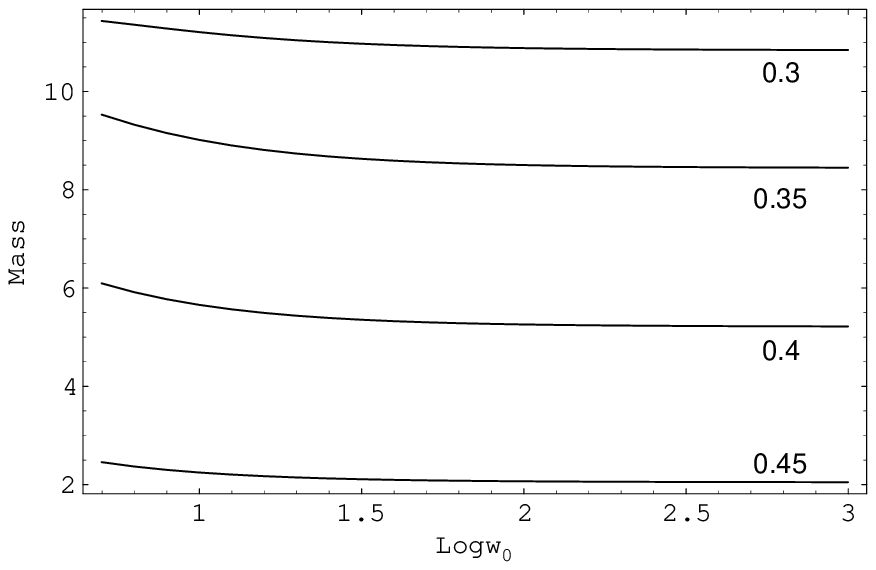}
\caption{The mass of a q-star as a function of $\omega_0$ for
four different values of $\omega$ in a scalar-tensor
gravitational theory with $n=1$.} \label{figure15}
\end{figure}

\begin{figure}
\centering
\includegraphics{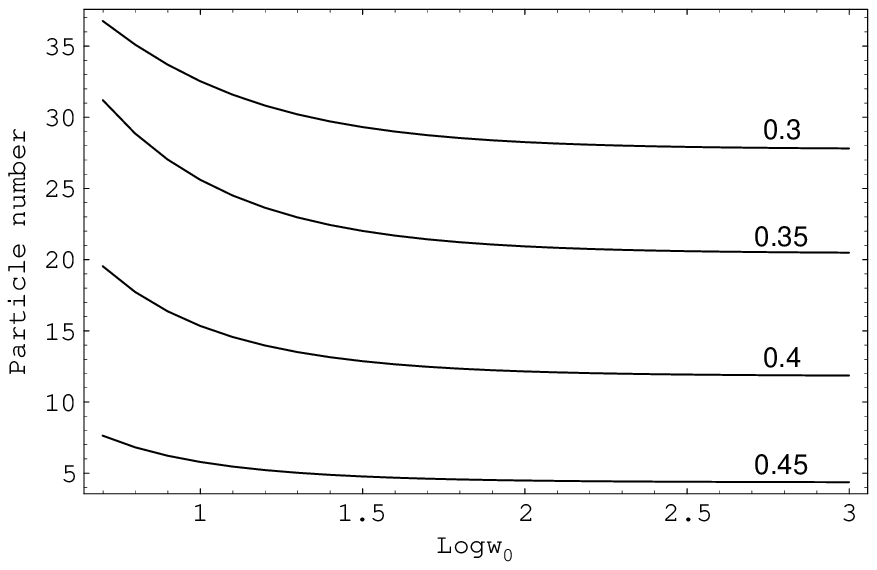}
\caption{The particle number of a q-star as a function of
$\omega_0$ for four different values of $\omega$ in a
scalar-tensor gravitational theory with $n=1$.} \label{figure16}
\end{figure}

We solved the coupled Einstein and and Lagrange equations for
several integral or half-integral values values of $n$ and found
that the star parameters are rather constant. This owes to the
$\Phi_{\textrm{BD}}\sim1$ relation, because $\omega_0$ and
$\omega_{\textrm{BD}}$ are not small enough so as to deviate
considerably from the results of General Relativity. When
$\omega_0$ decreases the star parameters are larger than in the
case of General Relativity, and when $\omega_0\rightarrow\infty$
its results are reproduced.

\section{Concluding remarks}

We investigated q-stars in a BD gravitational theory. We also
studied the case of q-stars in the framework of generalized scalar
tensor theories, with $\omega_{\textrm{BD}}$ a simple polynomial
function of the BD scalar. All the field configurations discussed
here are stable with respect to fission into free particles as the
ratio of their energy to the energy of the free particles, equal
to the mass of the free particles times the particle number, is
smaller than unity. We investigated their properties, particle
number, mass, radius of the matter field configuration and the
value of the matter scalar field at the center of the star. The
free parameters of their phase space are mainly the
eigenfrequency, straightforwardly connected with the surface
gravity, and the value of the $\omega_{\textrm{BD}}$ or the
$\omega_0$ for the generalized scalar-tensor theory.

We found that the star parameters, mass, particle number, radius
and absolute value at the center, are in generally larger when
$\omega_{\textrm{BD}}$ or $\omega_0$ is small and coincide with
the results of general relativity when
$\omega_{\textrm{BD}},\hspace{1em} \omega_{0} \rightarrow\infty$.

\vspace{1em}

\textbf{ACKNOWLEDGMENTS}

\vspace{1em}

I wish to thank N. D. Tracas and E. Papantonopoulos for helpful
discussions.

\end{document}